\documentclass[conference,9pt]{IEEEtran}
\usepackage[T1]{fontenc}

\usepackage{cite}
\usepackage{amsmath,amssymb,amsfonts}
\usepackage{graphicx}
\usepackage{booktabs}
\usepackage{algorithmic}
\usepackage{textcomp}
\usepackage{array}
\usepackage{tabularx}
\IEEEoverridecommandlockouts

\usepackage[hidelinks]{hyperref}

\hypersetup{
    pdfauthor={},
    pdfsubject={},
    pdfkeywords={}
}

\graphicspath{{figures/}}

\begin{document}
\raggedbottom

\title{ProcReverser: Multi-Agent Black-Box Microarchitectural Reverse
Engineering for Processor Vulnerability Analysis}

\author{
    \IEEEauthorblockN{
        Guanqi~Lyu\textsuperscript{1},
        Decheng~Qiu\textsuperscript{1},
        Guangyang~Zhang\textsuperscript{1},
        Fei~Zhao\textsuperscript{3},
        Zhaojun~Lu\textsuperscript{1},
        Jiaji~He\textsuperscript{1},
        and Yongqiang~Lyu\textsuperscript{1,2}
    }
    \IEEEauthorblockA{
        \textsuperscript{1}Tianjin University, Tianjin, China\\
        \textsuperscript{2}Tsinghua University, Beijing, China\\
        \textsuperscript{3}Peking University, Beijing, China
    }
    \thanks{This work has been submitted to the
    \MakeUppercase{ieee} for possible publication.
    Copyright may be transferred without notice,
    after which this version may no longer be accessible.}
}

\maketitle

\begin{abstract}
When processor implementation details are unavailable,
black-box microarchitectural reverse engineering helps
security researchers identify hardware conditions related
to security vulnerabilities. However, this process requires
substantial manual effort to design experiments, develop
test programs, and analyze results for each processor.
We present ProcReverser, a multi-agent system that
automates this process. ProcReverser uses execution
feedback to improve experimental plans and test programs
and identifies microarchitectural properties from
measurements. We develop a comprehensive suite of
property identification tasks focused on hardware
mechanisms relevant to Spectre-class attacks, covering
speculative execution, branch prediction, memory
dependence, and cache behavior. We evaluate ProcReverser
on the XuanTie C910, the gem5 RISC-V O3 CPU model,
and the SiFive P550. ProcReverser achieves Pass@1
and Pass@5 scores of 81.3\% and 93.8\%, respectively,
on C910, and 83.0\% and 100.0\% on gem5 O3.
On the SiFive P550, ProcReverser further characterizes microarchitectural
properties covered by the proposed task suite. These findings highlight
the potential of automated microarchitectural reverse
engineering to support processor vulnerability analysis
and defense evaluation.
\end{abstract}

\begin{IEEEkeywords}
microarchitectural reverse engineering, RISC-V, hardware security, multi-agent systems, large language model
\end{IEEEkeywords}

\section{Introduction}
\label{sec:introduction}

Assessing potential processor vulnerabilities requires understanding
how a target processor behaves under conditions relevant to
attacks and defenses. Even processors with the same instruction set architecture (ISA) may differ in branch prediction, speculative execution, and the handling of
microarchitectural state. These differences affect both the
conditions required for attacks and the effectiveness of
mitigations~\cite{gerlach2026riscv}. When implementation details are
unavailable, black-box microarchitectural reverse engineering uses
software experiments to infer hardware properties from observable
behavior~\cite{vila2020cachequery,chen2024branchpredictors}.

Such experiments often demand substantial manual
effort~\cite{liu2026ssbench}. Researchers must design tests for each
property, write and debug programs, establish controls, adapt
measurements to available processor interfaces, and interpret the
results. Moving tests to another processor may require changes to
assumptions, measurement methods, and decision
criteria~\cite{vila2020cachequery,chen2024branchpredictors}. Even
changes in a processor's internal configuration can affect the
results. Consequently, experiments often need to be redesigned and
validated for each target.

We present ProcReverser, a multi-agent system for black-box
processor microarchitectural reverse engineering. Given a property to identify and access to a target processor's execution
interface, its agents design and review experiments, generate test
programs, diagnose execution failures, and analyze measurements.
ProcReverser uses execution feedback to refine experimental plans
and programs, then infers properties from measured behavior. We also incorporate a mandatory evidence-checking mechanism, termed Discriminative Revalidation (DR). Each
reported conclusion includes supporting measurements and test
conditions, allowing researchers to examine its basis and scope.

Studies on RISC-V processor security have highlighted the importance of validating security properties on individual processor implementations. In 2018, the RISC-V Foundation stated that no RISC-V chips announced at the time were susceptible to Spectre or
Meltdown~\cite{riscv2018statement}. Spectre showed that instructions
executed on a mispredicted path can leave observable
microarchitectural effects even when their results are
discarded~\cite{kocher2019spectre}. More recently, Gerlach
et al.\ demonstrated multiple Spectre attacks on the XuanTie
C910/C920 and SiFive P550 and evaluated
mitigations~\cite{gerlach2026riscv}. They also noted that until a
dedicated speculation barrier extension with architectural guarantees
is standardized and supported in hardware, reliable Spectre
mitigations must be tested on each processor
implementation~\cite{gerlach2026riscv}. This need for
processor-specific testing motivates systematic identification of
the hardware properties that determine attack feasibility and
mitigation effectiveness. As RISC-V processors and their
configurations evolve, these properties and the validity of earlier
findings must be checked again.

The growing use of RISC-V in robotics, servers, and large language
model (LLM) inference further motivates such analysis. SpacemiT
reported its K3 chip in a humanoid
robot~\cite{spacemitk3,spacemit2026update}. V-Seek optimized LLM
inference on the server-class Sophon
SG2042~\cite{poveda2025vseek}. SiFive and AMD demonstrated LLM
inference on the BigSky development platform, with a RISC-V host
processor and GPU acceleration~\cite{sifive2026amd}. As RISC-V
platforms support increasingly complex workloads, identifying hardware properties relevant to processor vulnerabilities and checking mitigation effectiveness on real processors become
more important.

Motivated by the hardware conditions required for Spectre-class
attacks, we construct 26 security-relevant property identification
tasks covering speculative execution, branch prediction, memory
dependence, and cache behavior. We evaluate ProcReverser on two
commercial RISC-V processors, the XuanTie C910 and SiFive P550,
and the configurable gem5 RISC-V O3 CPU model. On C910, we check
identification correctness using public implementation information
and prior experimental findings~\cite{openc910,gerlach2026riscv}.
On gem5 O3, we vary internal settings while hiding them from the
agents to test adaptation to configuration changes. On P550, we
test whether ProcReverser can adapt its experiments to another
commercial processor and identify its microarchitectural
properties. Together, the three platforms allow us to assess
identification correctness, configuration adaptation, cross-processor
generalization, and the supporting experimental evidence.

Our main contributions are as follows.
\begin{enumerate}
    \item \textbf{Automated black-box microarchitectural reverse
    engineering.}
   We develop an experimental workflow in which five agent roles
   turn processor microarchitectural property questions into
   executable experiments and use measurements from target
   processors to produce evidence-backed property reports,
   reducing manual effort in microarchitectural reverse engineering.

    \item \textbf{A task suite for transient-execution security
    assessment.}
    We construct 26 property identification tasks covering
    speculative execution, branch prediction, memory dependence,
    and cache behavior, providing a common set of test questions for
    comparing processor implementations.

    \item \textbf{Cross-platform generalization.}
    We evaluate ProcReverser on the XuanTie C910, gem5 O3, and
    SiFive P550 to assess its identification accuracy and
    adaptability. The results demonstrate its ability to generalize
    across different processor implementations and adapt to varying
    microarchitectural configurations.
\end{enumerate}

\section{Related Work}
\label{sec:related-work}
\label{sec:background}

\subsection{Processor Security and Per-Implementation Validation}
\label{subsec:riscv-security}

Gerlach et al.\ show that Spectre attack conditions and mitigation
effectiveness depend on the implementation of commercial out-of-order
RISC-V processors, motivating validation on each target
processor~\cite{gerlach2026riscv}.
RISCover uses user-space differential testing to detect architectural
security vulnerabilities in closed-source RISC-V
processors~\cite{thomas2025riscover}.
WhisperFuzz uses RTL state transitions to detect timing vulnerabilities.
DejaVuzz combines dynamic swappable memory with differential
information-flow tracking to detect transient-execution
vulnerabilities~\cite{borkar2024whisperfuzz,xu2025dejavuzz}.

\subsection{Measurement-Based Microarchitectural Reverse Engineering}
\label{subsec:automated-testing}

uops.info, nanoBench, and PALMED provide instruction characterization,
low-overhead microbenchmarking, and runtime-based resource modeling,
respectively~\cite{abel2019uops,abel2020nanobench,derumigny2022palmed}.
Other studies infer the behavior of individual microarchitectural
mechanisms. CacheQuery learns cache replacement policies from hardware
measurements. SSBench automatically characterizes memory-dependence
predictors across architectures, including RISC-V. Chen et al.\
reverse-engineer conditional branch predictors in Apple Firestorm
and Qualcomm Oryon~\cite{vila2020cachequery,liu2026ssbench,
chen2024branchpredictors}.

\subsection{Automated Side-Channel Discovery and Black-Box Security Modeling}
\label{subsec:side-channel-modeling}

ABSynthe uses black-box experiments to synthesize side-channel attacks,
and Osiris generates and tests instruction sequences to discover timing
channels~\cite{gras2020absynthe,weber2021osiris}.
Revizor detects violations of speculation contracts, TREVEX identifies
anomalous transient data flows, and malcos synthesizes instruction-centric
leakage contracts from black-box observations~\cite{oleksenko2022revizor,
weber2026trevex,moreno2026malcos}.

\subsection{Agent-Assisted Hardware Experiments and Security Analysis}
\label{subsec:agent-automation}

uGen combines retrieval augmentation, multi-agent collaboration, and
execution feedback to generate and adapt microarchitectural attack
proofs of concept~\cite{dipta2026ugen}.
COVERT uses reference RTL and compilation and simulation feedback to
generate test programs for hardware Trojan detection~\cite{hasan2026covert}.
SCAgent applies LLM-assisted reasoning, verification, and data analysis
to operating-system side-channel analysis~\cite{xu2026scagent}.

Related efforts apply agents to hardware design and verification.
SV-LLM supports SoC security verification, MARC uses RTL and design
documentation for vulnerability detection and confirmation, and LASSO
uses formal-verification feedback to improve security-property and
assertion generation~\cite{saha2025svllm,rui2026marc,ankireddy2025lasso}.
Endukuri et al.\ combine ISA knowledge, RTL analysis, and execution-based
validation to generate RISC-V verification
stimuli~\cite{endukuri2026riscvstimulus}.
CHIA and Agentic Architect use agents for simulator alignment,
hardware research workflows, and microarchitectural design
optimization~\cite{cui2026chia,kypriotis2026agenticarchitect}.

These studies provide a methodological foundation for automated
hardware experiments. ProcReverser applies a multi-agent workflow
to black-box microarchitectural reverse engineering. It automates
experiment design, adaptation, and analysis to reduce the need for
sustained expert involvement in per-processor testing. We evaluate this
workflow on security-relevant property identification tasks across
RISC-V processors.

\section{Methodology}
\label{sec:methodology}

\subsection{Experimental Workflow}
\label{subsec:workflow}

\begin{figure}[!t]
    \centering
    \includegraphics[
        width=\columnwidth,
        height=68mm,
        keepaspectratio,
        pagebox=cropbox
    ]{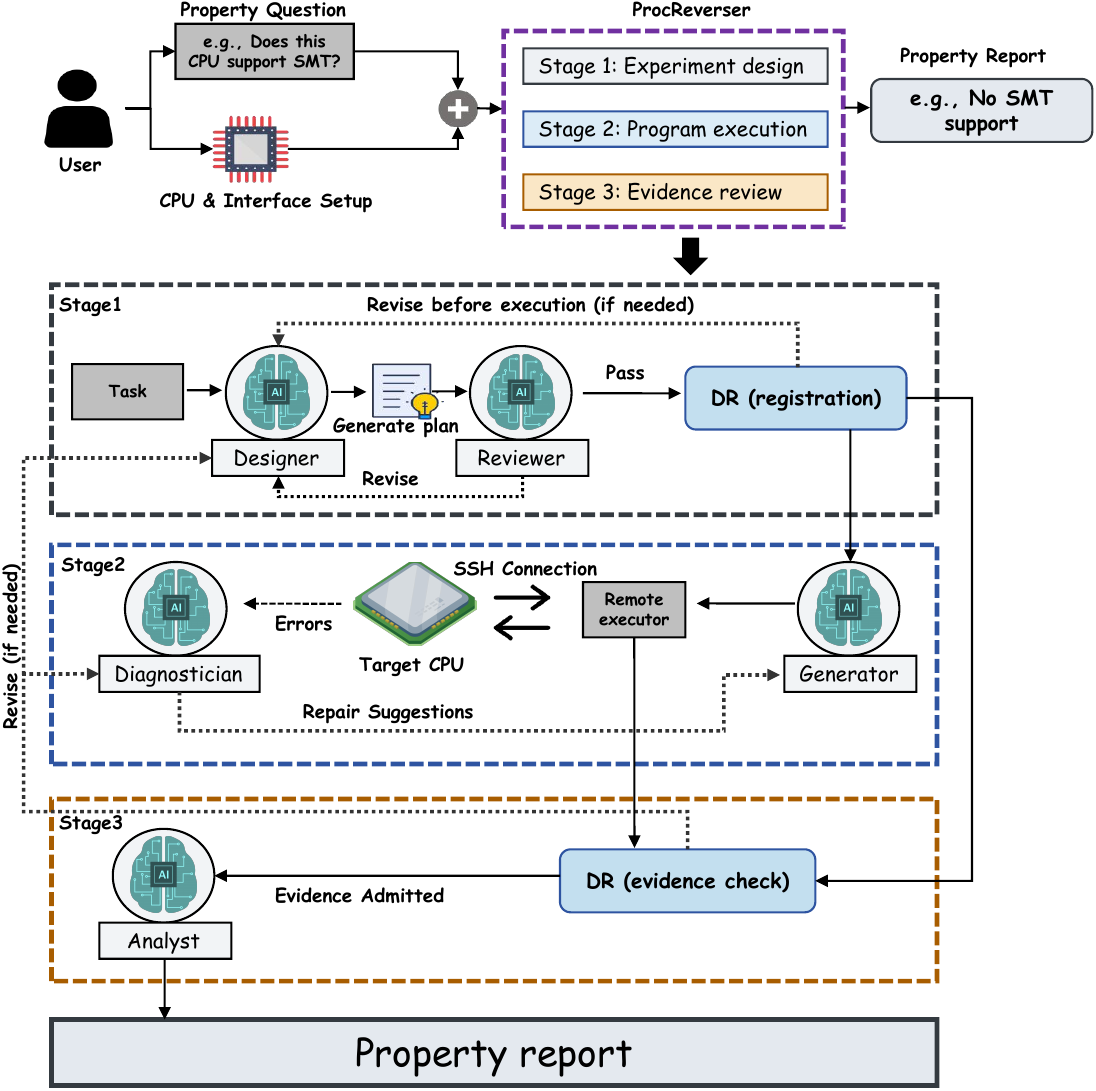}
    \caption{ProcReverser's experimental workflow. Solid arrows show the main workflow. Dashed arrows show feedback paths for error reporting, repair, and revision.}
    \label{fig:agent-workflow}
\end{figure}

As shown in Fig.~\ref{fig:agent-workflow}, the workflow has five
agent roles across three stages.

\textbf{Stage 1 (Experiment design).}
The \textbf{Designer} turns the property question into an
experimental plan with hypotheses, controls, and quantitative
decision criteria. The \textbf{Reviewer} checks whether the plan
is feasible on the target and can distinguish competing
explanations. Plans that need revision return to the Designer.
When DR is enabled, the approved plan is registered before execution.

\textbf{Stage 2 (Program execution).}
The \textbf{Generator} implements the plan, and the remote executor
runs the test on the target and returns measurements and execution
feedback. The \textbf{Diagnostician} examines the source code,
compiler messages, execution logs, and failed checks to recommend
repairs to the Generator.

\textbf{Stage 3 (Evidence review).}
When DR is enabled, it checks the raw measurements against the
registered plan before passing admitted evidence to the
\textbf{Analyst}. The Analyst then writes the property report.

The Designer may propose up to 10 primary plan versions per task
attempt. If the initial execution of a plan fails, the Generator
may repair and retry the program up to 10 times based on the
Diagnostician's recommendations.

\subsection{Task Definition and Property Set}
\label{subsec:tasks}

\begin{figure}[!t]
    \centering
    \includegraphics[width=\columnwidth]{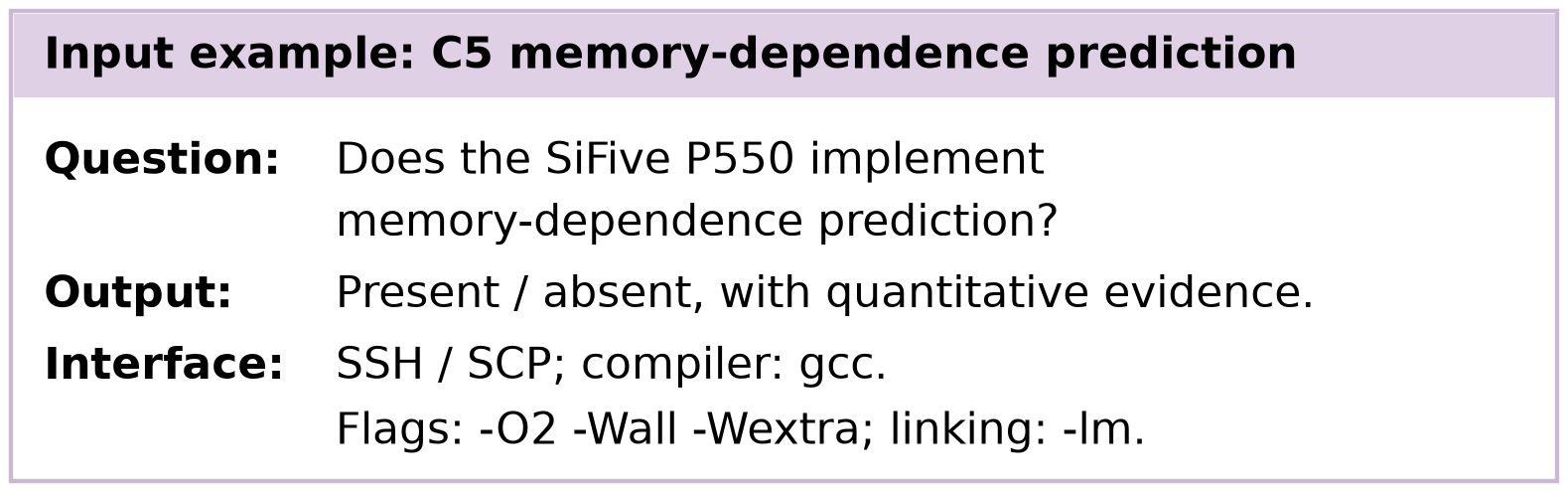}
    \caption{Example input for the C5 task.}
    \label{fig:task-input}
\end{figure}

ProcReverser takes as input a property question, the required output
fields, and the execution interface of the target processor. It outputs
a property statement and supporting
measurement evidence. Fig.~\ref{fig:task-input} shows an example input.

We construct the property set based on the key microarchitectural
conditions underlying major Spectre attacks. For Spectre-PHT,
Spectre-BTB, Spectre-RSB, and Spectre-STL, the set includes tasks that
measure conditional branch prediction, indirect branch target prediction,
return address prediction, and memory-dependence prediction and
speculative store bypass, respectively. The tasks also cover
cross-context state persistence, speculation windows, cache observation
conditions, and speculation-control mechanisms
\cite{kocher2019spectre,gerlach2026riscv,canella2019transient}.

These tasks cover speculative execution, branch prediction, memory
dependence, and cache behavior. They systematically characterize the
conditions needed to trigger attacks, influence microarchitectural state,
execute instructions transiently, and observe cache traces. They also
support analysis of mitigation applicability and measurement validity.

Table~\ref{tab:property-tasks} groups the 26 tasks by measurement target
and experimental purpose. Each task ID consists of a group letter
followed by a within-group index. Group \textbf{P} measures the effects
of instructions on speculative execution, group \textbf{R} measures
predictor behavior and structure, and group \textbf{W} measures
speculation windows. Group \textbf{C} examines observation interfaces,
caches, and memory accesses, group \textbf{F} examines predictor
enablement, and group \textbf{S} examines other transient phenomena and
whether software can clear predictor state.

\begin{table}[!t]
    \centering
    \caption{Security-relevant microarchitectural property tasks}
    \label{tab:property-tasks}
    \fontsize{7}{8}\selectfont
    \setlength{\tabcolsep}{1.5pt}
    \renewcommand{\arraystretch}{0.94}
    \begin{tabularx}{\columnwidth}{@{}X@{\hspace{5pt}}X@{}}
        \toprule
        {\begin{tabularx}{\linewidth}[t]{@{}l>{\raggedright\arraybackslash}X@{}}
            \textbf{ID} & \textbf{Property task} \\
            \midrule
            \textbf{P1} & Cache-signal suppression by tested instructions and transient-window type \\
            \textbf{P2} & Store-loop cycle overhead of tested barriers \\
            \textbf{P3} & Barrier effectiveness at tested instruction spacings \\
            \textbf{P4} & Cache-signal suppression from non-counter CSR reads \\
            \textbf{R1} & RAS capacity in return addresses \\
            \textbf{R2} & RAS overwrite on overflow and fallback when empty \\
            \textbf{R3} & Privilege-related predictor tagging and partitioning \\
            \textbf{R4} & Direction-predictor state persistence across address spaces or process switches \\
            \textbf{R5} & Direction-predictor state decay under time or branch interference \\
            \textbf{R6} & Branch-history dependence of direction prediction \\
            \textbf{R7} & BTB capacity by level or target configuration \\
            \textbf{R8} & Branch-history parameters and observable dependence \\
            \textbf{R9} & Predictor indexing, aliasing, and cross-context interference \\
        \end{tabularx}}
        & {\begin{tabularx}{\linewidth}[t]{@{}l>{\raggedright\arraybackslash}X@{}}
            \textbf{ID} & \textbf{Property task} \\
            \midrule
            \textbf{R10} & BTB entry insertion and update behavior \\
            \textbf{W1} & Transient-execution window length under tested triggers \\
            \textbf{W2} & Effects of tested instruction layouts on transient windows \\
            \textbf{C1} & Availability of user-mode timing and counter interfaces \\
            \textbf{C2} & Available observation channels and cache-maintenance operations \\
            \textbf{C3} & Cache capacity and organization of characterized levels \\
            \textbf{C4} & SMT support per physical core \\
            \textbf{C5} & Presence of a memory-dependence predictor \\
            \textbf{C6} & Hardware data prefetching enablement and observed behavior \\
            \textbf{C7} & Cross-core cache sharing (L2/LLC) \\
            \textbf{F1} & Enablement of tested branch-prediction components \\
            \textbf{S1} & Post-fault transient cache traces \\
            \textbf{S2} & Software control and observed clearing of predictor state \\
        \end{tabularx}} \\
        \bottomrule
    \end{tabularx}
    \par\vspace{2pt}
    \begin{minipage}{\columnwidth}
        \footnotesize\raggedright
        RAS (return address stack), BTB (branch target buffer),
        CSR (control and status register), SMT (simultaneous multithreading),
        LLC (last-level cache).
        \par\noindent
        \textit{P1 (P550).} The nine user-mode candidates are
        \texttt{fence} with \texttt{rw,rw}, \texttt{r,r},
        \texttt{r,w}, \texttt{w,r}, \texttt{w,w}, \texttt{i,i},
        or \texttt{iorw,iorw}, plus \texttt{fence.tso} and \texttt{fence.i}.
    \end{minipage}
\end{table}

\subsection{Discriminative Revalidation}
\label{subsec:dr}

During testing of the multi-agent system, we observed that,
once experiments yield analyzable data, the Analyst may adopt
an interpretation of the observations while overlooking the
controls or repeat measurements needed to distinguish
alternative causes. Without further validation, the system may therefore present a potentially incorrect conclusion based on insufficient evidence as definitive.
Recent studies have begun to enforce evidence requirements
within agent execution frameworks: AgInt uses its harness
to check evidence and control answer
submission~\cite{he2026agint}, whereas A-Harness triggers
targeted retries in response to evidence
gaps~\cite{huang2026aharness}.
Building on these evidence-guided agent workflows, we implement
\textbf{Discriminative Revalidation (DR)} within ProcReverser to determine
whether experimental measurements support the registered property hypotheses and to trigger targeted retests when evidence is insufficient.

DR is implemented in Python and integrated into the agent workflow. It consists of four steps.

\textbf{Step 1 (Registration).}
The Designer specifies candidate explanations, the comparisons
needed to distinguish them, the predicted outcomes, and the
quantitative decision criteria. After the Reviewer approves the
plan, DR registers these predictions and sends the plan back to
the Designer if different property answers have indistinguishable
predictions.

\textbf{Step 2 (Data comparison).}
After execution, DR computes the registered comparisons directly
from the raw samples in each batch and evaluates each batch
separately against the prespecified criteria.

\textbf{Step 3 (Targeted revision).}
If a required comparison is missing or a program-level check fails,
DR sends the specific gap to the Diagnostician, who recommends a
repair to the Generator. If the available evidence still supports
competing property answers or contradicts all registered
explanations, DR returns the unresolved comparisons to the Designer
for additional measurements or a revised experiment.

\textbf{Step 4 (Evidence admission).}
DR admits the evidence only when the validity checks pass, all
required comparisons yield clear outcomes, and all retained explanations imply the
same property answer. The Analyst uses the admitted evidence to
produce the property report.

\section{Experimental Setup}
\label{sec:experimental-setup}
\label{subsec:validation}

\textbf{Agent environment.}
ProcReverser's agents run on a local Windows 11 Pro computer with an
Intel Core Ultra 7 265K processor and 32 GB of RAM. We implement
ProcReverser using Python 3.13.5, LangChain 1.4.2, and LangGraph 1.2.12.
LangChain provides role-specific prompts, context management, and
tool calling. LangGraph coordinates role handoffs, conditional
branches, and iterative execution through a state graph
\cite{langchain_agents,langgraph_overview}.
All agents call Zhipu AI's GLM-5.3-Flash via its API, with
\texttt{reasoning\_effort} set to \texttt{high}
\cite{zai_glm53_flash}.

\textbf{Experimental platforms.}
We use two hardware development boards and one configurable
simulation platform. The Sipeed Lichee Pi 4A uses a T-Head
TH1520 SoC with four 64-bit RISC-V XuanTie C910 cores
\cite{sipeed_licheepi4a}. The SiFive HiFive Premier P550 uses
an ESWIN EIC7700X SoC with four 64-bit out-of-order SiFive P550
cores \cite{sifive_hifive_p550}. We use the RISC-V O3 CPU model
in gem5 25.1.0.1 and configure its out-of-order pipeline,
branch predictors, and cache system to construct simulated
targets that can run RISC-V test programs
\cite{lowepower2020gem5,gem5_o3cpu}.

\textbf{Remote execution and cluster environment.}
The local controller interacts with the target platforms through
a unified execution interface. For both development boards, the
executor connects to the onboard Linux system via SSH, uploads the
test program source code via SCP, compiles and runs the programs on
the board, and retrieves compiler output, exit statuses, execution
logs, and structured measurement results. For gem5, the executor
connects to the cluster login node via SSH, and Slurm 24.05.1
schedules simulation jobs on compute nodes. Each node used in the
experiments runs x86-64 Linux, has two 52-core processors, and is
configured in Slurm with approximately 1 TB of memory. The executor
uses Zig 0.16.0 on the cluster to cross-compile RISC-V test programs,
then runs gem5 and retrieves the measurement results. The agents use
execution feedback to revise experimental plans and test programs.

\textbf{Platform roles and evaluation settings.}
On C910, we compare inferred properties with publicly available
OpenC910 RTL details and prior empirical findings applicable to
the target implementation to verify structural parameters and
behavioral interpretations \cite{gerlach2026riscv,openc910}.
In gem5, we vary the target configuration and withhold configuration
files, implementation source code, and reference answers from the
agents. With the language model, agent workflow, and task definitions
fixed, we assess whether the system can adapt its experiments based
on execution feedback and whether its identification results track
changes in the target configuration. On P550, we assess whether the
system can adapt to different prediction behaviors, measurement
interfaces, and execution conditions and produce property reports
for commercial hardware. C910 and P550 use GCC 12 and GCC 13.3.0,
respectively, with the compilation flags \texttt{-O2 -Wall -Wextra}.
Each reference task receives five independent attempts.

\section{Results and Analysis}
\label{sec:evaluation}

\subsection{Reference Tasks and Scoring}
\label{subsec:reference-scoring}

To assess identification correctness against independent
references, we establish separate scoring sets for C910 and
gem5 RISC-V O3 from the common inventory of 26 tasks.
For C910, applicable public RTL, user manuals, prior empirical
studies, and board documentation provide references for
16 tasks. For O3, implementation source code and the actual
configurations of individual target instances provide
references for 20 tasks.
Tables~\ref{tab:c910-reference-answers}
and~\ref{tab:o3-reference-answers} summarize the scoring tasks
and reference answers for C910 and O3, respectively.

We exclude the remaining six gem5 O3 tasks from correctness
scoring because their reference answers cannot be established
from simulator configurations alone. Some require additional
controlled experiments to calibrate program- and
measurement-dependent outcomes. To ensure fair and reliable
evaluation, we score only tasks with independently
verifiable reference answers.

Each scoring task receives five independent attempts using the same
tools and model settings. Reference
answers and implementation information used for scoring are withheld
from the agents. An attempt passes if its final conclusion matches
the corresponding reference answer in Table~\ref{tab:c910-reference-answers}
for C910 or Table~\ref{tab:o3-reference-answers} for gem5 O3.

Let $N$ denote the number of scoring tasks for a platform,
and let $c_i$ denote the number of successful attempts among
the five for task $i$. We compute
\begin{equation}
    \mathrm{Pass@1}
    = \frac{\sum_{i=1}^{N} c_i}{5N},
    \qquad
    \mathrm{Pass@5}
    = \frac{1}{N}
      \sum_{i=1}^{N}\mathbf{1}[c_i>0].
    \label{eq:pass-metrics}
\end{equation}
For C910 and O3, $N$ is 16 and 20, respectively.
Metrics are calculated separately for each platform.

\begin{table}[!t]
    \centering
    \setlength{\abovecaptionskip}{2pt}
    \caption{C910 reference answers.}
    \label{tab:c910-reference-answers}
    \footnotesize
    \setlength{\tabcolsep}{2.5pt}
    \renewcommand{\arraystretch}{0.88}
    \begin{tabularx}{\columnwidth}{@{}l>{\raggedright\arraybackslash}X@{}}
        \toprule
        \textbf{ID} & \textbf{Independent reference answer} \\
        \midrule
        \textbf{P1} & \textbf{Load/store fault tests.} Read-ordering fences, \texttt{fence.i}, and counter CSR reads suppress the cache signal. \texttt{fence w,w}, \texttt{fence.tso}, and \texttt{lr.d} do not. [C] \\
        \addlinespace[1.5pt]
        \textbf{R1} & RAS capacity: \textbf{12 entries}. [A,B,C] \\
        \textbf{R2} & \textbf{Overflow:} circular overwrite. \textbf{Underflow tests:} no BTB fallback or straight-line speculation. [B,C] \\
        \textbf{R3} & \textbf{RAS/Ind-BTB:} privilege tags. \textbf{BHT:} no explicit ASID/privilege index key. [B] \\
        \textbf{R4} & PHT training \textbf{crosses address spaces}. [C] \\
        \textbf{R6} & \textbf{Branch history} contributes to direction indexing. [B] \\
        \textbf{R7} & L0BTB: \textbf{16 entries}. [B] \\
        \textbf{R8} & \textbf{22-bit global history} with slicing and XOR folding. [B] \\
        \textbf{R9} & \textbf{Core-local history} in finite index mappings. PHT training crosses address spaces. [B,C] \\
       \textbf{R10} & \textbf{BTB:} refill-buffer updates,
with \texttt{tag[2:1]} selecting write enables. [B] \\
        \addlinespace[1.5pt]
        \textbf{W1} & \textbf{Window (instructions):} null-pointer fault~$\sim$60, immediate bounds~$\sim$7, delayed bounds/indices~$\sim$61. [C] \\
        \addlinespace[1.5pt]
        \textbf{C2} & User-mode \textbf{cache flushing and Flush+Reload} are available. [C] \\
        \textbf{C3} & \textbf{L1I/L1D:} each 64\,KiB, two ways, 64\,B lines. L1I is VIPT with FIFO updates. L1D is PIPT. [A,B,D] \\
        \textbf{C4} & \textbf{No SMT}. [C] \\
        \textbf{C7} & Four cores with private L1I/L1D and a \textbf{shared 1\,MiB L2}. [D] \\
        \addlinespace[1.5pt]
        \textbf{S2} & \textbf{M-mode MCOR} (\texttt{0x7c2}) controls invalidation of BHT, Ind-BTB, and BTB (including L0BTB). [A,B] \\
        \bottomrule
    \end{tabularx}
    \par\smallskip
    \begin{minipage}{\columnwidth}
        \footnotesize\raggedright
        \textit{Sources.} [A] User manual (June 27, 2024)~\cite{c910_user_manual_2024}.
        [B] OpenC910 RTL
        \href{https://github.com/XUANTIE-RV/openc910/tree/b91c90914c19f114d35c8f6b73408eb241ed847c}{\texttt{b91c909}}~\cite{openc910}.
        [C] Gerlach et al.\ (measurements)~\cite{gerlach2026riscv}.
        [D] Sipeed~\cite{sipeed_licheepi4a}.
    \end{minipage}
\end{table}

\begin{table}[!tb]
    \centering
    \setlength{\abovecaptionskip}{2pt}
    \caption{Reference answers for 20 gem5 O3 tasks.}
    \label{tab:o3-reference-answers}
    \footnotesize
    \setlength{\tabcolsep}{2.5pt}
    \renewcommand{\arraystretch}{0.89}
    \begin{tabularx}{\columnwidth}{@{}l>{\raggedright\arraybackslash}X@{}}
        \toprule
        \textbf{ID} & \textbf{Reference configuration and behavior} \\
        \midrule
        \textbf{P4} & \texttt{fcsr}, \texttt{fflags}, and \texttt{frm} reads \textbf{suppress the cache signal} from wrong-path loads in the tested SE branch window. \\
        \addlinespace[1.5pt]
        \textbf{R1} & RAS: \textbf{8 entries} in the baseline, \textbf{16} in the alternate instance. \\
        \textbf{R2} & \textbf{Overflow:} overwrites the oldest entry. \textbf{Underflow:} stale address or empty slot. If empty, keep the BTB target if present, else fall through. \\
        \textbf{R3} & LocalBP has \textbf{no partitioning by privilege level}. \\
        \textbf{R4} & LocalBP state \textbf{persists across same-core process switches}. \\
        \textbf{R5} & LocalBP counters \textbf{do not decay with time alone}. \\
        \textbf{R6} & LocalBP indexes by \textbf{PC alone}. \\
        \textbf{R7} & BTB: \textbf{512 entries in SE}, \textbf{4096 in FS}. \\
        \textbf{R8} & TournamentBP uses \textbf{13 history index bits} (\texttt{0x1FFF}). \\
        \textbf{R9} & \textbf{LocalBP:} same index, shared counter. \textbf{BTB:} a new tag at the same index replaces the old entry. \\
        \textbf{R10} & \textbf{BTB updates} on eligible taken-branch corrections. \\
        \addlinespace[1.5pt]
        \textbf{C1} & \textbf{SE:} counter CSRs and clocks work. \textbf{FS:} \texttt{rdcycle} traps. \texttt{rdtime} and \texttt{clock\_gettime} work. \\
        \textbf{C2} & \texttt{cbo.flush} executes at valid addresses in \textbf{SE and FS}. \\
        \textbf{C3} & L1D: \textbf{64\,KiB, four ways, 64\,B lines}. \\
        \textbf{C4} & \textbf{One hardware thread} per core. \\
        \textbf{C5} & \textbf{StoreSet} memory-dependence prediction is present. \\
        \textbf{C6} & \textbf{Prefetch triggers off} in the baseline. No requests issued. \\
        \textbf{C7} & Two cores have private L1s and a \textbf{shared 2\,MiB L2}. \\
        \addlinespace[1.5pt]
        \textbf{F1} & \textbf{Enabled:} direction, direct/indirect target, and return prediction. \\
        \textbf{S2} & \textbf{No full clearing of trained direction state observed} after \texttt{fence.i} (1/8/64 times), \texttt{fence rw,rw}, or \texttt{nop} in SE/FS. In FS, \texttt{yield} and \texttt{sleep} caused partial disturbance. \\
        \bottomrule
    \end{tabularx}
    \par\smallskip
    \begin{minipage}{\columnwidth}
        \footnotesize\raggedright
        Note. SE (syscall emulation), FS (full-system simulation).
    \end{minipage}
\end{table}

\subsection{Identification Correctness}
\label{subsec:identification-correctness}

\begin{figure}[!t]
    \centering
    \includegraphics[width=\columnwidth]
        {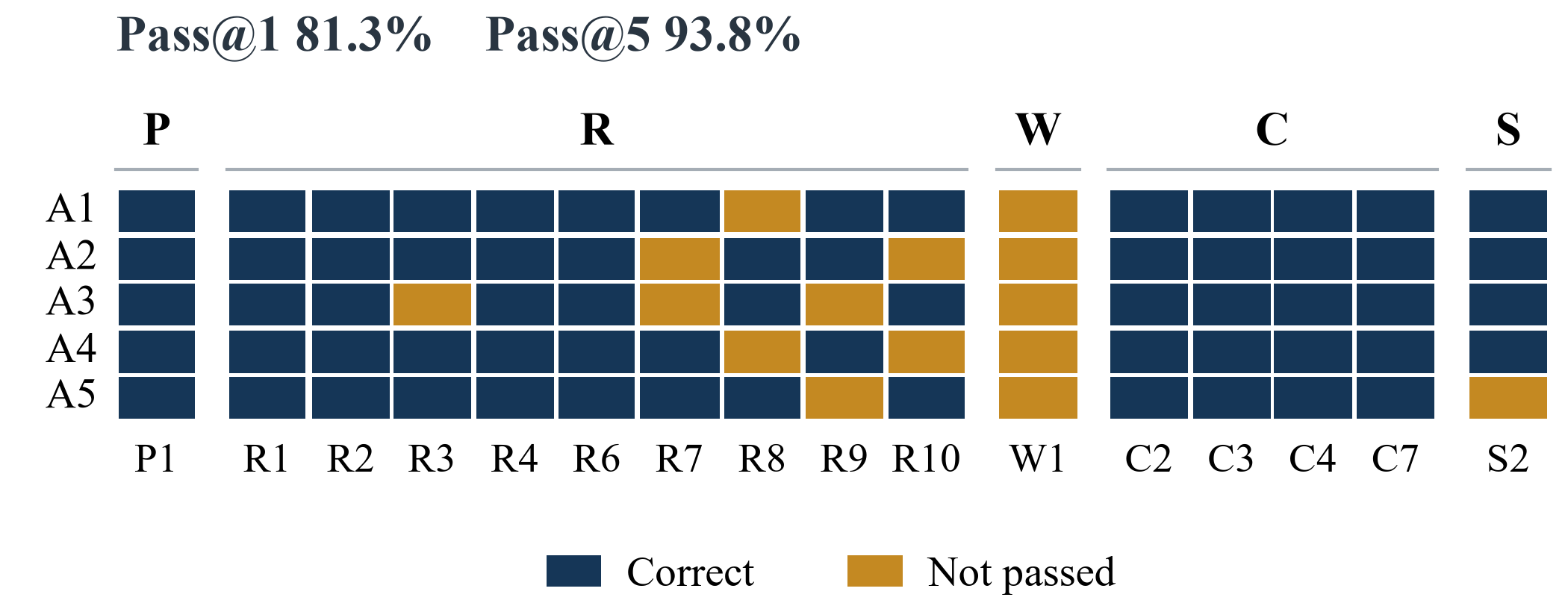}
    \caption{Identification outcomes for the 16 C910
    reference tasks. A1--A5 denote five independent attempts;
    each cell represents one scored task attempt.}
    \label{fig:c910-identification-matrix}
\end{figure}

\begin{figure}[!t]
    \centering
    \includegraphics[width=\columnwidth]
        {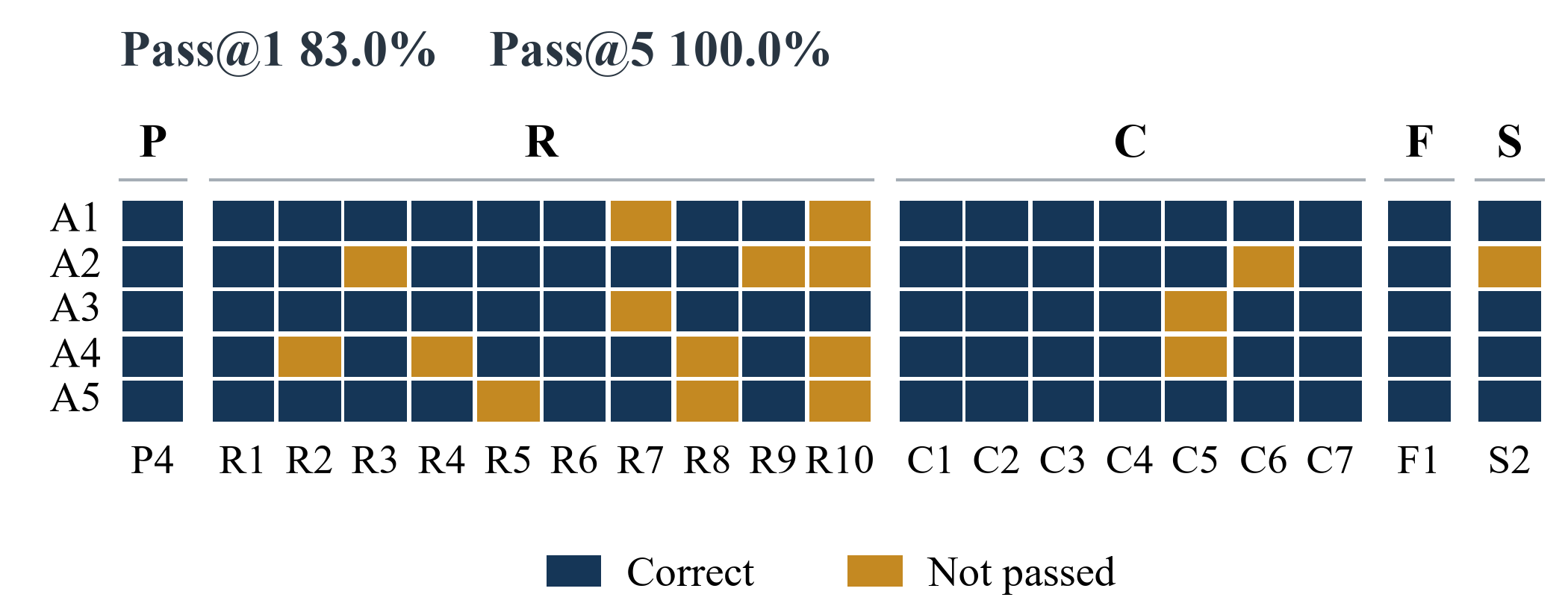}
    \caption{Identification outcomes for the 20 gem5 O3
    reference tasks. A1--A5 denote five independent attempts; each cell represents one scored task attempt.}
    \label{fig:o3-identification-matrix}
\end{figure}

Figs.~\ref{fig:c910-identification-matrix}
and~\ref{fig:o3-identification-matrix} show five independent
attempts on the C910 and gem5 O3 reference task sets,
respectively.
Each column represents a task, and each row represents
an independent attempt.
Dark blue indicates an outcome satisfying the success criterion
in Section~\ref{subsec:reference-scoring}, whereas gold
indicates an outcome that does not pass scoring.

For C910, 65 of the 80 attempts across 16 tasks pass scoring,
yielding a Pass@1 of \textbf{81.3\%}.
Fifteen tasks succeed at least once within five attempts,
yielding a Pass@5 of \textbf{93.8\%}.
For O3, 83 of the 100 attempts across 20 tasks pass scoring,
yielding a Pass@1 of \textbf{83.0\%}.
All 20 tasks succeed at least once, yielding a Pass@5
of \textbf{100.0\%}.
These metrics describe performance on the respective
platform-specific reference task sets.

The per-task outcomes further reveal differences in
identification stability.
On C910, W1 fails in all five attempts, whereas R7--R10
each succeed in three attempts.
On O3, R10 succeeds in only one attempt.
Repeated attempts improve task coverage, but some
predictor and speculation-window tasks still exhibit
repeated failures.

\subsection{Mechanism Ablation}
\label{subsec:mechanism-ablation}

We evaluate four configurations by running ProcReverser's five-role
workflow and a single-agent baseline, each with and without DR. The single agent performs
all duties in one context, while the five-role system distributes
them among the Designer, Reviewer, Generator, Diagnostician, and
Analyst. All configurations use the same initial information,
measurement tools, output requirements, and limits on primary
plan versions and repair retries specified in Section~\ref{subsec:workflow}. 

We impose no token budget limits, and all attempts terminate normally upon completing their respective
workflows and return results.
When enabled, DR follows the same procedure in both systems. We perform five independent attempts for each of the 16 C910 tasks and 20 gem5 O3 tasks and apply the
scoring criteria in Section~\ref{subsec:reference-scoring}.
Table~\ref{tab:mechanism-ablation} reports the results.

\begin{table}[!t]
    \centering
    \caption{Ablation of role specialization and DR.}
    \label{tab:mechanism-ablation}
    \footnotesize
    \setlength{\tabcolsep}{2pt}
    \renewcommand{\arraystretch}{1.05}
    \resizebox{\columnwidth}{!}{%
        \begin{tabular}{@{}lrrrrrrr@{}}
            \toprule
            Configuration
            & \multicolumn{2}{c}{C910}
            & \multicolumn{2}{c}{O3}
            & \multicolumn{3}{c}{Mean cost} \\
            \cmidrule(lr){2-3}
            \cmidrule(lr){4-5}
            \cmidrule(lr){6-8}
            & Pass@1 & Pass@5 & Pass@1 & Pass@5
            & Plans & Exec. & Tokens (M) \\
            \midrule
            Single agent
            & 17.5\% & 37.5\% & 18.0\% & 40.0\%
            & 7.06 & 75.09 & 3.34 \\
            Single agent + DR
            & 22.5\% & 50.0\% & 25.0\% & 45.0\%
            & 7.83 & 47.66 & 2.87 \\
            Five roles
            & 68.8\% & 87.5\% & 72.0\% & 90.0\%
            & 4.44 & 40.36 & 13.11 \\
            Five roles + DR
            & \textbf{81.3\%} & \textbf{93.8\%}
            & \textbf{83.0\%} & \textbf{100.0\%}
            & 5.66 & 22.55 & 12.88 \\
            \bottomrule
        \end{tabular}%
    }
\par\vspace{2pt}
\begin{minipage}{\columnwidth}
    \scriptsize\raggedright
    \textit{Note.} Means are computed over all task attempts
    per configuration across C910 and gem5 O3. Exec.\ counts program
    execution attempts. Tokens (M) counts input and output tokens
    combined, in millions.
\end{minipage}
\end{table}

\textbf{Effect of role specialization.}
With and without DR, the five-role configuration achieves higher
Pass@1 and Pass@5 than the corresponding single-agent configuration
on both C910 and gem5 O3. Without DR, Pass@1 rises from 17.5\% to
68.8\% on C910 and from 18.0\% to 72.0\% on O3. Across all task
attempts on both platforms, the five-role configurations use fewer
plan versions on average and make 46.3\%--52.7\% fewer program
execution attempts, while consuming approximately 3.9--4.5 times as
many tokens as their single-agent counterparts. The gains in
identification success and target-side execution come at a higher
model inference cost.

\textbf{Effect of DR.}
Enabling DR increased the observed Pass@1 and Pass@5 for both
agent organizations on both platforms. In the five-role
configuration, Pass@1 increased by 12.5 percentage points on
C910 (95\% CI [6.25, 21.25]) and by 11.0 percentage points on
gem5 O3 (95\% CI [4.00, 18.00]). We computed 95\% confidence intervals using 10,000
paired bootstrap resamples of tasks within each platform,
preserving the results with and without DR for each
resampled task.

Across both platforms, enabling DR was associated with
36.5\% and 44.1\% fewer mean program execution attempts
for the single-agent and five-role configurations,
respectively, without increasing their mean
token consumption. This may reflect fewer Generator and Diagnostician iterations.

\textbf{Case study of DR.}
We use the C6 hardware prefetching task in gem5 O3 to show how
DR identifies a missing comparison, directs a targeted retest,
and checks whether the resulting evidence supports a conclusion. The Designer specified two measurements: comparing
the cost of sequential and shuffled scans, and checking whether
a probe cache line left untouched during training was already
cached at its first access. The first version of the program
returned only scan timings. The sequential and shuffled scans
took 6.91 and 15.78 cycles per cache line, respectively.
With DR disabled, ProcReverser ran three rounds of diagnosis and
program repair to check whether the timing gap persisted.
It increased the number of scans, adjusted the working-set size,
and changed the shuffle seed in successive rounds. All three
rounds returned only scan timings. The Analyst ultimately
misinterpreted the persistent timing gap as evidence of hardware
prefetching.

With DR enabled, DR checked the raw samples against each
preregistered comparison before the Analyst reviewed the results.
The first run returned scan timings for both conditions but lacked
samples from the first access to the probe line after sequential
and shuffled training. The DR report identified the missing
comparison, and the workflow routed the task to diagnosis.
The Generator then added the required probe sampling. In the
discovery phase, each condition was tested in 200 trials, yielding
three cache hits after sequential training and two after shuffled
training. A separate confirmation batch yielded two and three
hits, respectively. Both batches met the preregistered criterion
for no detection, which required a probe hit rate of at most
5\% in each condition. The Analyst submitted the conclusion that
no hardware prefetching was detected under the tested access
patterns.

\subsection{Configuration Adaptation and P550 Property Identification}
\label{subsec:adaptation-characterization}

To examine ProcReverser's adaptability across configurations
and processors, we present two case studies: RAS depth
identification under two concealed gem5 O3 configurations
and hardware prefetching characterization on SiFive P550.
We further summarize the results of all 26 property
tasks on P550.

\begin{figure}[!tb]
    \centering
    \includegraphics[width=\columnwidth]{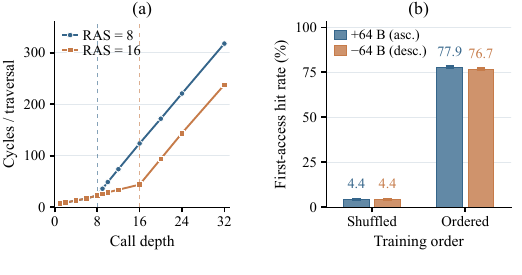}
    \caption{(a) RAS depth identification on gem5 O3.
    (b)  Hardware prefetching characterization on P550. Error bars indicate 95\% confidence
intervals for the mean hit rates.}
    \label{fig:adaptation-p550}
\end{figure}

\begin{table}[!tb]
    \centering
    \caption{Property characterization results on SiFive P550.}
    \label{tab:p550-characterization}
    \footnotesize
    \setlength{\tabcolsep}{2.5pt}
    \renewcommand{\arraystretch}{1.0}
    \begin{tabularx}{\columnwidth}{@{}l>{\raggedright\arraybackslash}X@{}}
        \toprule
        \textbf{ID} & \textbf{Measured conclusion} \\
        \midrule
        \textbf{P1} & All nine user-mode barriers listed in Table~\ref{tab:property-tasks} showed attenuation of the measured cache signal in both memory-dependence and fault-triggered windows. \\
\textbf{P2} & In a store loop, each execution of \texttt{fence r,w},
\texttt{fence w,w}, and \texttt{fence.i} added about 3, 10, and
46 cycles, respectively. \texttt{fence rw,rw}, \texttt{fence r,r},
\texttt{fence w,r}, and \texttt{fence.tso} each added about 13 cycles. \\
        \textbf{P3} & \texttt{fence.i} suppresses the fault-window cache channel at tested spacings of 0--16 instructions. \\
        \textbf{P4} & \texttt{fcsr} reads do not suppress the tested fault window. \\
        \addlinespace[1pt]
        \textbf{R1} & 16-entry RAS. \\
        \textbf{R4} & Same-core direction-prediction state persists across processes. \\
      \textbf{R5} & No half-decay with up to 256 sleep calls (200~$\mu$s requested each) or 1.024M interfering branches. \\
        \textbf{R6} & Prediction uses history at tested distances up to 128 branches. \\
        \textbf{R8} & Observed history-dependence window: 168--183 branch distances. \\
        \textbf{R9} & Histories distinguishable at the same PC. Cross-process same-PC training interferes. \\
        \textbf{R10} & Mispredictions fall after repeated execution. \\
        \addlinespace[1pt]
        \textbf{W1} & Branch window is about one instruction when the branch operands are immediately available. \\
        \textbf{W2} & No reproducible large layout effect with 0, 4, 8, 16, 32, 44, 60, or 128 bytes of leading padding. \\
        \addlinespace[1pt]
        \textbf{C1} & \texttt{perf} counts cycles. User-mode \texttt{rdcycle}/\texttt{rdinstret} trap. \\
        \textbf{C2} & L1 set-conflict, direction-prediction, data-TLB, and cross-core last-level cache channels available. User-mode \texttt{cbo.flush} traps. \\
        \textbf{C3} & L1D $\sim$32\,KiB. Shared last-level cache $\sim$4\,MiB. \\
        \textbf{C4} & No SMT. \\
        \textbf{C5} & Memory-dependence prediction present. \\
        \textbf{C6} & Bidirectional prefetching detected. \\
        \textbf{C7} & Four cores share the last-level cache. L2 sharing undetermined. \\
        \textbf{F1} & Conditional branch direction prediction was enabled. \\
        \addlinespace[1pt]
        \textbf{S1} & Post-fault loads leave observable transient cache traces. \\
        \textbf{S2} & Tested \texttt{fence.i}, branch flooding, and sleep-based switches do not restore cold predictor state. \\
        \multicolumn{2}{@{}l@{}}{\textbf{Undetermined:} R2, R3, R7.} \\
        \bottomrule
    \end{tabularx}
\end{table}

\textbf{Return address stack identification under configuration changes.}
The agent system autonomously designed a nested-call experiment with
multiple call sites, progressively increasing the call depth and measuring
the cost of returns. As shown in Fig.~\ref{fig:adaptation-p550}(a), the
change point in return cost shifts to a greater call depth as RAS capacity
increases. Two-segment linear fits place the change points at approximately
7.65 and 15.75, yielding capacity estimates of 8 and 16 entries,
respectively, consistent with the configured capacities of the two gem5 O3
instances.

\textbf{Hardware prefetching identification.}
ProcReverser designed and conducted a first-access experiment on the
SiFive P550 at strides of \(+64\) and \(-64\)\,B. Ordered and
shuffled training accessed the same cache lines and ended on the
same line. The system then timed the first access to a candidate
line that had not been demand-accessed during training. It set a
153-cycle hit threshold using hot- and cold-cache measurements
from separate calibration runs. Each of the 20 confirmation
batches included 512 trials per direction and training condition.
Across the two directions, mean hit rates were 99.54\%--99.66\%
for hot-cache controls and at most 0.01\% for cold-cache controls.

Fig.~\ref{fig:adaptation-p550}(b) shows mean first-access hit rates
of 77.9\% (ascending) and 76.7\% (descending) after ordered
training, versus 4.4\% in both directions after shuffled training.
The respective differences were 73.5 and 72.3 percentage points
(95\% CIs [72.6, 74.3] and [71.3, 73.2]), supporting bidirectional
hardware prefetching on the P550.

Table~\ref{tab:p550-characterization} summarizes the results of 26 property tasks on P550.
ProcReverser reports property conclusions for 23 tasks, while R2,
R3, and R7 remain undetermined due to insufficient evidence.
Some conclusions are partial, covering the subproperties
supported by the available measurements.
These results characterize the processor's observable
microarchitectural behavior and provide references
for assessing side-channel attack conditions and evaluating
potential defenses.

\section{Discussion and Limitations}
\label{sec:discussion-limitations}

\textbf{Limitations.}
Our evaluation is limited to RISC-V and covers two commercial
processors, the XuanTie C910 and SiFive P550, as well as the
configurable gem5 RISC-V O3 CPU model. The results show that
ProcReverser adapts to different RISC-V implementations and
configurations. Its applicability to processors implementing other
ISAs, such as Arm and x86, remains to be evaluated. Some properties may remain undetermined when target-side measurement signals are weak
or when the results provide insufficient evidence to distinguish
competing explanations. All experimental configurations use the same
language model. Controlled experiments are needed to assess how model
choice affects experimental design and property identification
accuracy. The ablation study also shows that agent role specialization
reduces the number of target-side program executions and increases
model inference costs.

\textbf{Practical applications.}
When processor implementation details are unavailable, ProcReverser
can use black-box measurements to test hardware conditions relevant
to transient-execution risks. It integrates experiment design,
test program generation, failure diagnosis, and evidence analysis
into a single workflow, reducing the need for sustained expert
involvement in per-processor testing. Each property report documents
the conclusion, supporting measurements, and test conditions.
These reports help researchers decide which hardware properties warrant
further validation and under what conditions to evaluate mitigations,
thereby providing an experimental basis for processor vulnerability
analysis.

\textbf{Future work.}
We plan to evaluate ProcReverser's generalization on more commercial
processors, including different microarchitectural implementations
of the same ISA and processors implementing other ISAs such as Arm
and x86. We will also explore new agent design methods for
microarchitectural reverse engineering and compare the performance of
different language models and agent collaboration strategies.

\section{Conclusion}
\label{sec:conclusion}

We presented ProcReverser, a multi-agent system for black-box
microarchitectural reverse engineering to support processor
vulnerability analysis. It uses execution feedback to refine
experiments and Discriminative Revalidation (DR) to enforce
evidence checks and trigger targeted retests when evidence is
insufficient, improving identification accuracy.
ProcReverser achieves Pass@1 and Pass@5 scores of 81.3\%
and 93.8\% on the XuanTie C910, and 83.0\% and 100.0\%
on gem5 RISC-V O3, respectively. On the SiFive P550,
the system identifies microarchitectural properties in
23 of the 26 tasks.
These results show that ProcReverser can identify
security-relevant microarchitectural properties across
the evaluated processors and configurations while providing
supporting measurements and experimental conditions.
These capabilities provide an experimental basis for
analyzing the hardware conditions underlying Spectre-class
vulnerabilities, assessing the applicability of mitigations,
and supporting further processor vulnerability analysis.

\end{document}